\documentclass[aps,epsfig,preprint]{revtex4}
\usepackage[section]{placeins}        
\usepackage{float}
\usepackage{kpfonts}
 \usepackage{graphics}
 \usepackage{multirow}
\usepackage{epsfig}
\usepackage{pdfpages}
\usepackage{amsmath,amsthm, amssymb, latexsym}
\usepackage{color}
\usepackage{graphics}
\usepackage{epsfig}
\usepackage[outercaption]{sidecap}
\DeclareMathAlphabet{\mathcal}{OMS}{cmsy}{m}{n}
\SetMathAlphabet{\mathcal}{bold}{OMS}{cmsy}{b}{n}

\newcommand{\ts}{\textstyle}

\newcommand{\gsim}{\raisebox{-0.6ex}{\mbox{ $\stackrel{\ts >}{\ts \sim}$ }}}

\newcommand{\bee}{\begin{equation}}
\newcommand{\ene}{\end{equation}}
\newcommand{\beea}{\begin{eqnarray}}
\newcommand{\enea}{\end{eqnarray}}

\begin{document}
\title{ Drift waves with dust acoustic wave coupling}
\author{Atul Kumar$^1$}
\email{atul.j1211@gmail.com}
\author{Amita Das$^2$}
\email{amitadas3@yahoo.com}
\author{Predhiman Kaw$^{1,0}$}

\affiliation{$^1$Institute for Plasma Research, HBNI, Bhat, Gandhinagar - 382428, India,\\
$^2$ Department of Physics, Indian Institute of Technology, Delhi, Hauz Khas, New Delhi-110016, India}
\begin{abstract}  
Drift wave is a prominent mode of a magnetized plasma of inhomogeneous density. It plays an important role in the transport of particles, energy and momentum perpendicular to the ambient magnetic field. 
The frequency of this mode is governed by the inhomogeneity scale length and is much lower 
than the typical homogeneous plasma modes involving ions and electrons. 
 In this work the 
 possibility of coupling of this particular mode  with the  low frequency modes of a dusty plasma medium 
 is  considered.  
 

\end{abstract}
\maketitle 
\section{ Introduction}

\par\vspace{\baselineskip}
The study of low frequency drift modes ( frequency much lower than   
 ion cyclotron frequency )  has received  widespread attention \cite{BB:1993,BB:1996,BB:19961,BB:68,BB:77, BB:78}.  These   waves are  excited in the presence of  plasma density 
inhomogeneity in a magnetized plasma and  are responsible for transport in such a system \cite{BB:1999}.  In this work 
we seek a possible coupling of this low frequency mode with the modes  of a dusty plasma medium \cite{BB:01,BB:02,BB:09,BB:931,BB:991,BB:031,BB:962,BB:2015,BB:2012}. 
As is well known the 
dusty plasma has,  in addition to electron and ions,  a third heavier 
dust species which typically gets negatively charged by the attachment of electron species 
from the background plasma. The 
  presence of third heavier dust species  in plasma leads to additional low frequencies 
  modes in the system. These low frequency modes of the dusty plasma medium 
  can couple with the low frequency drift wave modes under appropriate parameter regime.  
  This paper explores such a possibility.

\footnotetext[0]{Predhiman Kaw is a deceased author.}  
A classical Coulomb plasma with micron sized dust particles can lie in strongly coupled regime  \cite{BB:86} because the coupling parameter $ \Gamma = \frac{(Z_d e)^2}{4\pi \epsilon_0 aT_d} $ can easily be of the order $1$ or larger ($ Z_d e $ is the charge on the dust grain, $a $ is the inter-grain distance and $ T_d $ is the dust temperature). The dust species in the plasma acquires a high charge ($Q \sim 1000 e$) where $e$ is the electronic charge and has a  very low temperature ($T_d \sim 0.1 eV $) due to which it can be found in a strongly coupled regime. 
Recent experiments also confirm that dust particles can form crystal structures \cite{BB:04PRL,BB:94PRL}. There are many approximate models to represent the dynamics of strongly coupled fluid. One such model is Generalised Hydrodynamic (GHD)  model, first reported by Berkovsky \cite{BB:921} for the electron-ion plasma where ion fluid was treated as strongly coupled. This model was later adopted by Kaw and Sen \cite{BB:98} in dusty plasma to study low frequency mode characteristics when dust particles are strongly coupled. This model predicted the  existence of transverse shear mode in strongly coupled dusty plasma which has been later confirmed in many simulations\cite{BB:142,BB:14} as well as experiments\cite{BB:071}.  The GHD model will be adopted by us here to depict the dynamics of 
dust species. 


The manuscript has been organized as follows. Section II describes a model set of equations for the coupling of drift waves to the dust acoustic wave. 
In section III, a dispersion relation has been obtained. The dust species 
is treated both as a weakly and strongly correlated fluid system. 
In section IV we numerically plot the  dispersion relation and discuss the effect 
of coupling of the drift wave with the dust species.  
Section V contains the summary.

\section{Model equations for drift waves in dusty plasma}
\begin{center}
  \begin{figure}[!h]
 \centering
 \fbox{\includegraphics[width=0.5\textwidth]{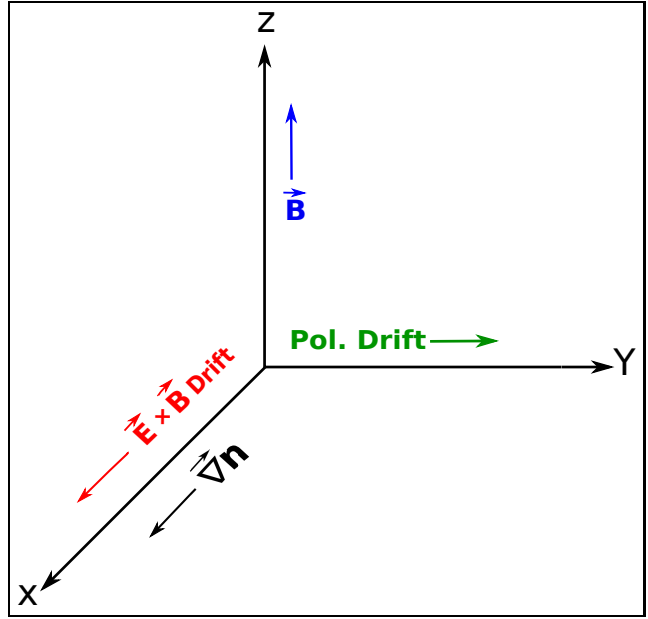}}
 \caption{Schematics of system configuration for the coupling of drift wave with dust acoustic wave }
\label{fig1}
 \end{figure}
 \end{center}
 We consider the plasma medium in a slab geometry,  schematics has been depicted in Fig.1. The 
 applied external magnetic field is $\vec{B}$ is directed along $\hat{z}$ axis. The magnetic field strength 
 is such that while the lighter electron and ion species are magnetized, the dust species remains 
 unmagnetized.  The plasma density 
 inhomogeneity $\nabla n$
 is chosen to be  along $\hat{x}$. 
 The suffix $ s =i,e,d $ refers to the   ion, electron and dust species of the plasma  respectively.  
 All the three species  are assumed to have  an equilibrium density profile which is inhomogeneous 
 and has an  exponential  form  given by the expression:
\begin{equation}
{n_{0s}(x)  = n_{00s} \exp\left(-\frac{x}{L_{n}} \right) }
\label{denexpr}
\end{equation}
In equilibrium the charge  balance is provided by the condition. 
\begin{equation}
n_{0i}=n_{0e}+Z_dn_{0d}
\label{quasineutrality}
\end{equation}
For drift wave like perturbations it is assumed that the wavelength perpendicular to the magnetic field 
is much smaller than the parallel wavelength. The  lighter electron species provides a 
balance of electric and pressure forces in the parallel direction to yield the following 
   Boltzmann relationship for the electron density perturbations:
\begin{equation}
{n_{e1} = n_{e0} \exp \left( \frac{e\phi}{T_e} \right) }
\label{eleBoltzeq}
\end{equation}
 Here, $ T_e$  corresponds to the  electron temperature and $ \phi $ is the electrostatic potential. 
 Similarly, we denote ion and dust temperatures by $T_i$ and $T_d$ respectively. 
 

The  ion continuity equation is 
\begin{equation}
\frac{\partial n_i}{\partial t} + \vec{\nabla}_\perp (n_iv_{i\perp}) + \vec{\nabla}_z (n_iv_{iz})=0
\label{ioncontnuity}
\end{equation}
The perpendicular ion momentum equation yields $v_{i\perp}$ the ion drift velocity. 
Using the low frequency drift approximation wherein $\omega/\omega_{ci} \ll 1$ 
we retain the $\vec{E} \times \vec{B}$ and polarization drift velocities for the ions 
\begin{equation}
{ \vec{v}_{i\bot} = \frac{ \hat{z} \times \vec{\nabla} \phi }{B} + \frac{1}{ \omega_{ci}B} \bigg[ - \frac{ \partial }{ \partial t } {\nabla _{\perp}^2} \phi - 
  \frac{ \hat{z} \times \vec{\nabla} \phi }{B}\cdot \vec{\nabla} \vec{\nabla} \phi \bigg]   }
\label{ionperpvel}
\end{equation}
Here the first term $\vec{V}_E= \hat{z} \times \vec{\nabla}\phi/B$ is the $\vec{E} \times \vec{B}$ drift velocity of ions and the second term in the polarization drift velocity $\vec{V_p} $ of the ions. 
 We now  substitute the expression of $ \vec{v}_{i\bot} $ in ion continuity equation and employ the  quasineutrality condition of  $Z_d n_{d1} + n_{e1} = n_{i1}  $ to eliminate 
 ion density perturbation. Here $ n_{d1} $, $ n_{e1} $, $ n_{i1} $ are the perturbed densities for dust, electrons and ions respectively. 
 Furthermore, the electron density perturbation $ n_{e1} $, is replaced by the scalar potential  
  using the linearized Boltzman  relation provided by Eq.(\ref{eleBoltzeq}).  One thus obtains
\begin{equation}
 \frac{\partial}{\partial_t} \bigg( n_{e0} \frac{e \phi}{T_e} + Z_dn_{d1} \bigg) + n_{i0} \bigg[ - \vec{\nabla}_\bot \cdot \bigg\lbrace \frac{1}{ \omega _{ci} B} \frac{ \partial }{ \partial t } ( \vec{\nabla}_\bot \phi ) \bigg\rbrace - \vec{\nabla}_\bot \bigg\lbrace \frac{1}{ \omega_{ci}B} \vec{V}_E \cdot \vec{\nabla}( {\nabla_ {\perp}^2} \phi ) \bigg\rbrace \bigg] + \vec{V}_E \cdot \vec{\nabla}n_{i0} + n_{i0} \vec{\nabla} \cdot \vec{v}_{iz}= 0 
\end{equation}
Here the smallness of  polarization drift $ \vec{V}_P $  by $\omega/\omega_{ci}$  compared to  $ \vec{E} \times \vec{B} $ has been considered. 
Rearranging the above equation one obtains an expression as:
\begin{equation}
{ Z_d \frac{\partial}{\partial_t}(n_{d1}) +  \frac{\partial}{\partial_t} \big[n_{e0} \frac{e \phi}{T_e} - \frac{n_{i0}}{ \omega_{ci}B}{\nabla_ {\perp}^2} \phi \big] + \frac{ \hat{z} \times \vec{\nabla} \phi }{B} \cdot \vec{\nabla}\bigg[ ln(n_{i0})-\frac{1}{ \omega_{ci}B}{\nabla_ {\perp}^2} \phi \bigg] + n_{i0} \vec{\nabla} \cdot\vec{v}_{iz}= 0 }
\label{MHW}
\end{equation} 
Eq.(\ref{MHW}) is a modified equation for the drift wave dynamics in the presence of dust species.  
The last term represents  the ion motion in the direction parallel to the ambient magnetic field which is 
often  ignored in the usual drift wave treatment as variations parallel to the magnetic field is 
assumed to be much smaller. 
The first term in Eq.(\ref{MHW})   is dependent on the dust density and represents the coupling 
with the dust dynamics.  The ion dynamics in the  in the direction parallel to the ambient magnetic field is written as:
\begin{equation}
\frac{d \vec{v}_{iz}}{dt} = -\frac{e}{m_i}  \vec{\nabla} \phi 
\label{parmom}
\end{equation}
where $k_z \ll k_\perp$ has been considered.

The dust density is provided by the continuity equation:
\begin{equation}
\frac{\partial}{\partial_t}(n_{d}) +  \vec{\nabla}\cdot (\vec{v}_d n_d)= 0
\label{dustcont}
\end{equation}
The dust velocity is determined by the momentum equation for which we use the generalized 
hydrodynamic equation representing the visco-elastic behaviour of the dust fluid 
in the strong coupling limit  
\begin{equation}
\bigg [ 1 + \tau _m \frac{d}{dt} \bigg] \bigg[\frac{d \vec{v}_d}{dt} -Z_d e n_{d0} \vec{\nabla} \phi \bigg ] = \eta {\nabla}^2 \vec{v}_d
\label{GHDM}
\end{equation}
The parameter  $\tau_m$ represent the elastic relaxation time corresponding to  the 
memory associated with the  strong coupling limit and $\eta$ is the viscosity parameter. 
It can be observed that for a physical phenomenon lying in the frequency domain of 
$\omega \tau_m \ll 1$ the dust fluid behaves like an ordinary hydrodynamic fluid. However, viscoelastic effects associated with strong coupling manifest in the limit of $\omega \tau_m \gsim 1$. 
It should also be noted that we have considered the strength of applied magnetic field 
such that the dust response remains unmagnetized. 
We thus have Eqs.(\ref{MHW},\ref{dustcont},\ref{GHDM}) as the complete set of equations 
representing the coupling of drift wave with the dust acoustic waves. 

In the next section, we obtain the linear dispersion relation of the above set of equations in both  
limits of weakly and strongly coupled response of the dust fluid.

\section{Linear dispersion relation of drift wave coupled with dust acoustic wave}
Linearizing Eq.(\ref{MHW}) yields
\begin{equation}
Z_dn_{d1}+ \bigg [ n_{e0} \frac{e \phi}{T_e} + \frac{n_{i0}}{\omega _{ci}B}{k_{\perp}^2} \phi \bigg] + \frac{k_y \phi}{\omega B} \vec{\nabla}(n_{i0}) - \frac{n_{i0}e{k_{z}v_{iz}} }{\omega} =0
\label{linMHM}
\end{equation}
Here $k_\perp^2 = k_x^2+k_y^2$ is the perpendicular wave number and $k_x,k_y$ are the wavenumbers in the $\hat{x}$ and $\hat{y}$ directions respectively. 
The ion momentum equation (Eq.(\ref{parmom})) in parallel direction can be linearized as:
\begin{equation}
v_{iz} = \frac{ek_z \phi}{m_i\omega}
\label{linparmom}
\end{equation}
where $k_z$ is the parallel wavenumber such that the wavenumber in the system can be given as $k^2=k_\perp^2+k_z^2$.

Similarly the linearization of the dust continuity and momentum equation leads to 
\begin{equation}
n_{d1} = n_{d0} \frac{k}{\omega}v_{d1}
\label{dustcont_lin}
\end{equation}
\begin{equation}
\bigg(\omega+\frac{\iota\eta k^2}{m_d n_{d0}(1-\iota\omega\tau_m)}\bigg)v_{d1} = -\frac{Z_dn_{d0}k\phi}{m_d\omega}
\label{GHDM_lin}
\end{equation}
The dust velocity $v_{d1}$ is directed along the electric field $\vec{E} =-\nabla \phi$  which has no component along  $\hat{x}$ the inhomogeneity direction. Thus the term 
$\vec{v_{d1}} \cdot \nabla n_{0d} $ has not been included in the linearized continuity equation for 
the dust species. Combining the equations (\ref{linMHM}-\ref{GHDM_lin}) leads to the following dispersion relation:\begin{equation}
-{\bigg[{\omega}^2+  \frac{\iota \omega\eta k^2 }{m_dn_{d0}(1-\iota \omega \tau_m)} \bigg]}^{-1}{C_D^2k^2} +\bigg [ \frac{n_{e0}}{n_{i0}} + {k_{\perp}^2} {\rho _s^2} \bigg] +\frac{k_y T_e}{\omega eB} \vec{\nabla}(ln[n_{i0}]) - \frac{{k_z^2}{v_s^2}}{{\omega}^2}=0
\label{fulldisp}
\end{equation}

Here $ \omega _{pd}= \sqrt{\frac{4\pi {Z_d^2}e^2n_{d0}}{m_d}} $, $ \lambda _{di}=\sqrt{\frac{T_i}{4 \pi n_{i0}e^2 }} $ are the plasma frequency of dust particles and ion Debye length, $ \rho_s = \sqrt{ \frac{T_e}{m_i} } \cdot \frac{1}{ \omega_{ci} } $ is the effective ion Larmor radius at the electron temperature $T_e$ and $ v_s = \sqrt{\frac{T_e}{m_i}} $ is the acoustic speed of ions. The dust acoustic velocity $ C_D $ is given by $ C_D = \omega _{pd} \lambda _{di} $. \\

Using the normalization $ \frac{\omega}{\omega _{ci}} \rightarrow \omega $, $ k \rho _s \rightarrow k $, $ \omega _{ci}t \rightarrow t $, $ \frac{x,y}{\rho _s} \rightarrow x,y $,$\frac{L_n}{\rho _s} \rightarrow L_n $, $ \omega _{ci} \tau_m \rightarrow \tau_m $, $ \frac{\eta}{m_dn_{do}\omega_{ci}{\rho_s^2}} \rightarrow \eta $, above dispersion relation expressed in Eq.(\ref{fulldisp}) can be  obtain  as,

\begin{equation}
-{\bigg[{\omega}^2+  \frac{\iota \omega\eta k^2 }{(1-\iota \omega \tau_m)} \bigg]}^{-1}{\frac{{C_D^2}}{{v_s^2}}k^2} +\bigg [ \frac{n_{e0}}{n_{i0}} + {k_{\perp}^2}  \bigg] -\frac{1}{L_n} \frac{k_y}{\omega}  - \frac{{k_z^2}}{{\omega}^2}=0
\label{disprel}
\end{equation}

This expression is the dispersion relation for the ion drift wave in presence of dust dynamics. Now Eq.(\ref{disprel}) can be analysed in its  two asymptotic limits namely: (1)Hydrodynamic limit \emph{i.e} $ \omega \tau_m \ll 1 $ in which dust particles are weakly correlated to each other  and, (2) Kinetic limit $ \omega \tau _m \gg 1 $, in which dust particles are strongly correlated. 
\subsection{Hydrodynamic limit, $ \mathbf{\omega \tau_m \ll 1} $}
In the limit where $ \omega \tau_m  \ll 1 $ which is the case for weakly coupled dust particles, Eq.(\ref{disprel}) can be written as,
\begin{equation}
-\frac{{C_D^2}}{{v_s^2}}\frac{k^2}{{({\omega}^2+  {\iota \omega\eta k^2 } })} +\bigg [ \frac{n_{e0}}{n_{i0}} + {k_{\perp}^2} {\rho _s^2} \bigg] -\frac{1}{L_n} \frac{k_y}{\omega}  - \frac{{k_z^2}}{{\omega}^2}=0
\label{MHMWD}
\end{equation}
Eq.(\ref{MHMWD}) is the modified Hasegawa-Mima equation in weakly coupled dusty plasma. The presence of viscosity term  $\eta k^2$ would lead to the  damping of drift wave coupled to dust acoustic wave. However, in the low viscosity limit \emph{i.e.} $\omega \gg \eta k^2 $, the above equation (Eq.\ref{MHMWD}) reduces to a simple quadratic equation as
\begin{equation}
\bigg ( \frac{n_{e0}}{n_{i0}} + {k_{\perp}^2}\bigg ) {\omega}^2 -\omega_* \omega - \bigg [ \frac{{C_D^2}}{{v_s^2}} k^2 + {k_z^2} \bigg] =0 
\end{equation}
where $ \omega _* = k_y/L_n $ is the classical drift wave frequency. The roots of the above equation are
\begin{equation}
 \omega = \frac{{\omega_* } \pm \sqrt{{\omega_*^2}+ 4 \bigg( \frac{n_{e0}}{n_{i0}} + {k_\perp^2} \bigg ) \bigg ( \frac{{C_D^2} {k}^2}{{v_s^2}} + {k_z^2} \bigg ) }}{2\bigg( \frac{n_{e0}}{n_{i0}} + {k_\perp^2} \bigg )} 
 \label{WCDroots}
\end{equation}
The variation in $k_x$ is assumed to be very slow such that the wavelength in $\hat{x}$-direction to be very small compared to the scale length of the density gradient so that we can still linearize in the $\hat{x}$-direction in spite of the inhomogeneity in that direction.

\subsection{Kinetic limit, $ \mathbf{\omega \tau_m \gg 1} $}
The  dust particles being at low temperature and having high charge, can readily go in the strongly coupled regime, while electrons and ions having comparatively very high temperature, can still be considered as light fluid. The dynamics of strongly coupled dust is no longer similar to that described by the ordinary fluid in the hydrodynamic limit $ \omega \tau_m \ll 1 $. Now its dynamics resembles to that shown by viscoelstic  fluid. In the kinetic regime \emph{i.e} $ \omega \tau_m \gg 1 $, Eq.(\ref{disprel}) can be written as,

\begin{equation}
-\frac{{C_D^2}}{{v_s^2}}\frac{k^2}{(\omega ^2-\frac{\eta}{\tau_m} k^2)} +\bigg [ \frac{n_{e0}}{n_{i0}} + {k_{\perp}^2}  \bigg] -\frac{1}{L_n} \frac{k_y}{\omega}  - \frac{{k_z^2}}{{\omega}^2}=0
\end{equation}
This equation can be further reduced to a fourth order equation in $ \omega $ as,
\begin{equation}
\bigg [ \frac{n_{e0}}{n_{i0}} + {k_{\perp}^2} \bigg ] {\omega}^4 - \frac{k_y}{L_n}{\omega}^3 - \bigg [ \frac{{C_D^2}}{{v_s^2}}k^2 + \bigg ( \frac{n_{e0}}{n_{i0}} + { k_{\perp}^2} \bigg ) \frac{\eta}{\tau _m} k^2 + {k_z^2} \bigg ]{ \omega}^2 + \frac{k_y}{L_n} \frac{\eta}{\tau _m} k^2 \omega + \frac{\eta}{\tau_m} {k_z^2}k^2 =0
\label{MHMSD}
\end{equation}
Eq.(\ref{MHMSD}) is the modified Hasegawa-Mima equation in strongly coupled dusty plasma.

\section{ Results and  analysis}
The coupling of drift waves with dust acoustic wave happens when the condition $ \omega \sim \omega_* \sim kC_D \sim k_z v_s $ is satisfied. 
The frequency $\omega= \omega_r + \iota \gamma$ can be complex where $\omega_r$ and $\gamma$ are  the real and imaginary part. It has been shown that the  coupling of drift wave with dust acoustic wave does not lead to instability. However, the presence of dust dynamics has been shown to strongly modify the dispersion properties of drift waves. 
\subsection{Weakly correlated dusty plasma ($ \mathbf{\omega \tau_m \ll 1} $)}
 It is evident from the roots [Eq.(\ref{WCDroots})] that there is a finite value of $ \omega_r $ even at $ k_y =0 $. Also we observe that mode saturates to $\omega_r = 1\times 10^{-4}$ the large $ k_y $ region where the mode frequency is independent of $ k_y $ [see Eq.(\ref{WCDroots})]. It is due to the  shielding by  polarization drift in the perpendicular direction. The dispersion relation is valid in the regime where $ {k_\perp^2} {\rho_s^2} \gg 1 $ because we have taken $ T_i \ll T_e $. Therefore even if $ {k_\perp^2} {\rho_i^2} \ll 1 $ (condition for ions to be magnetized, where $ \rho_i=\frac{1}{\omega_{ci}} \sqrt{\frac{T_i}{m_i}}$ is the ion Larmor radius at ion temperature), we can have the condition $ {k_\perp^2} {\rho_s^2} \gg 1 $  valid.

\begin{center}
  \begin{figure}[!h]
 \centering
 \fbox{\includegraphics[width=0.7\textwidth]{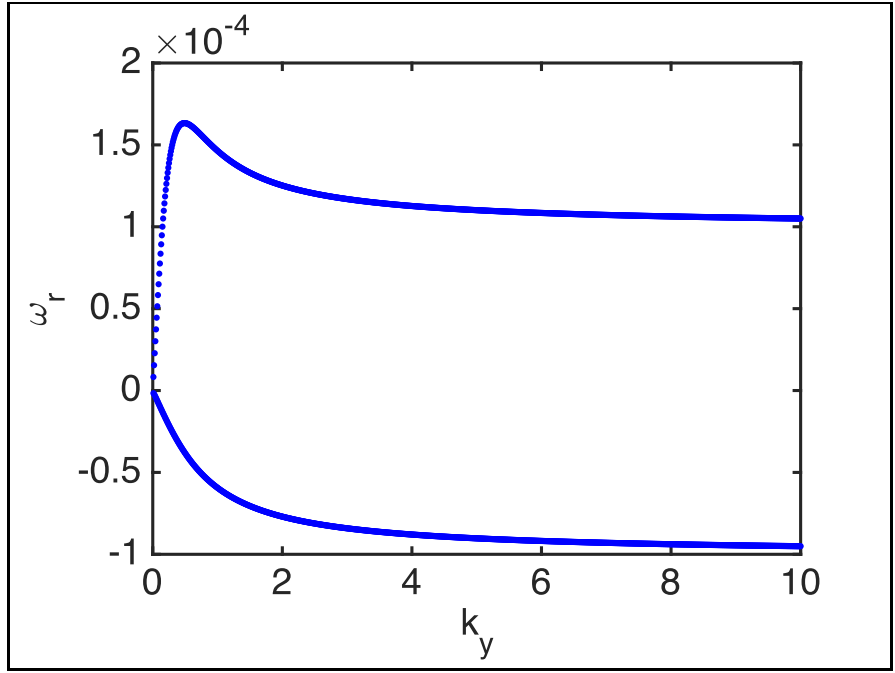}}
 \caption{Dispersion Relation  ($ k_z=10^{-6} $, $ L_n=10^4 $, $ \frac{n_{e0}}{n_i0} = 0.15 $, $ \frac{C_D}{v_s} = 10^{-4} $ ) where we show that the mode saturates to $1$ for large value of $k_y$ }
\label{fig2}
 \end{figure}
 \end{center}
 \begin{center}
  \begin{figure}[!h]
 \centering
 \fbox{\includegraphics[width=0.7\textwidth]{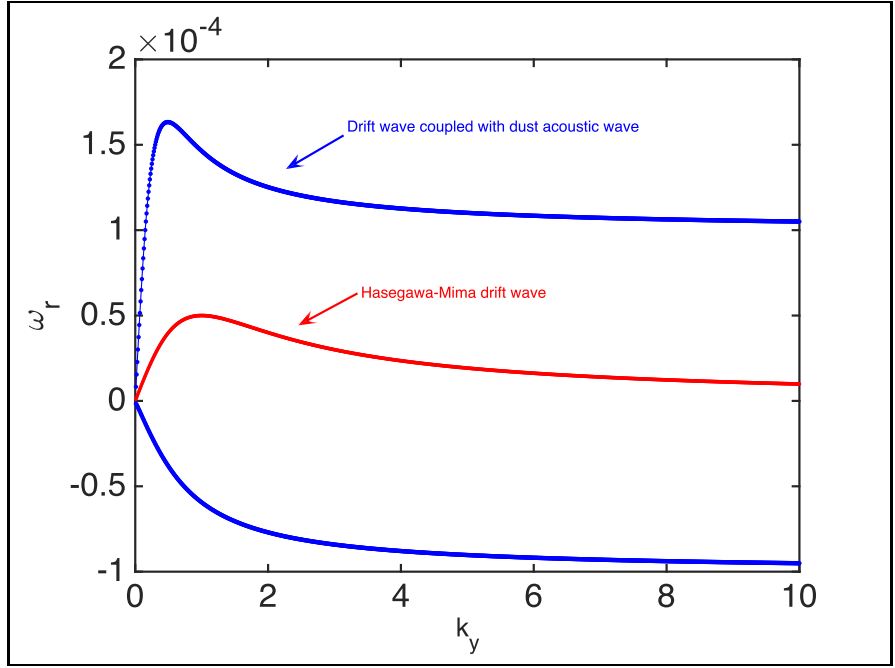}}
\caption{Comparison of dispersion relations between Hasegawa-Mima drift wave and the drift wave coupled with dust acoustic wave for parameters ($ k_z=10^{-6} $, $ L_n=10^4 $, $ \frac{n_{e0}}{n_i0} = 0.15 $, $ \frac{C_D}{v_s} = 10^{-4} $ )}
\label{fig3}
 \end{figure}
 \end{center}
 \begin{center}
  \begin{figure}[!h]
 \centering
 \fbox{\includegraphics[width=0.7\textwidth]{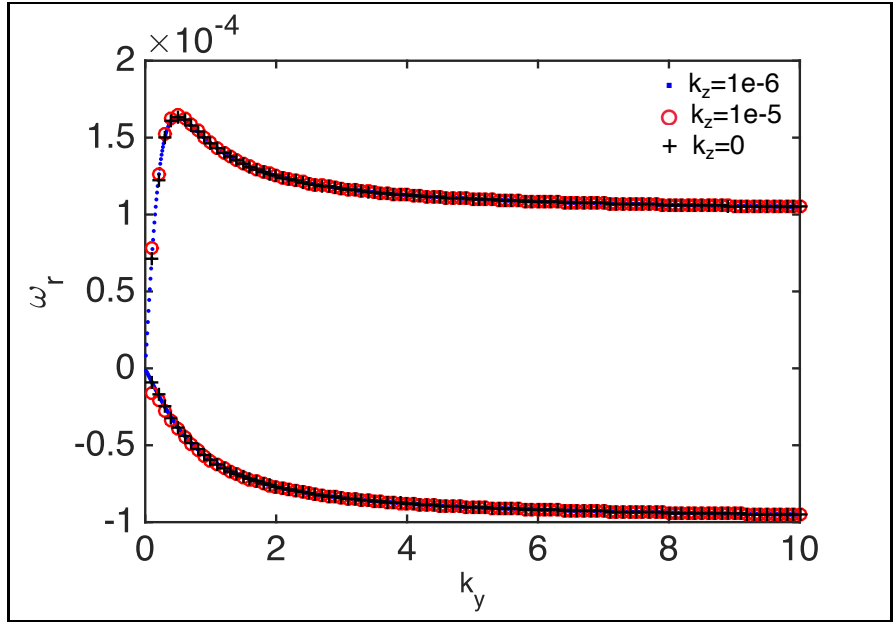}}
\caption{The effect of $k_z$ on the dispersion relation of the drift wave coupled to dust acoustic wave ( $ L_n=10^4 $, $ \frac{n_{e0}}{n_i0} = 0.15 $, $ \frac{C_D}{v_s} = 10^{-4} $)}
\label{fig4}
 \end{figure}
 \end{center}
 \begin{center}
  \begin{figure}[!h]
 \centering
 \fbox{\includegraphics[width=0.7\textwidth]{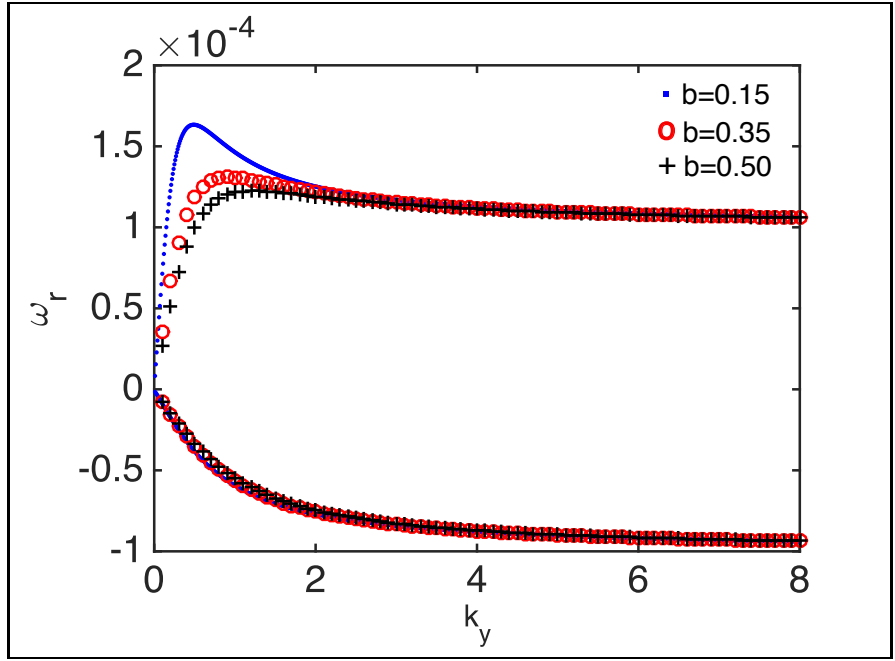}}
 \caption{The effect of depletion ratio of electrons $b=n_{e0}/n_{i0}$ on the dispersion relation of the drift wave coupled to dust acoustic wave ( $ L_n=10^4 $, $ k_z=10^{-6}$, $ \frac{C_D}{v_s} = 10^{-4} $)}
\label{fig5}
 \end{figure}
 \end{center}
 \begin{center}
  \begin{figure}[!h]
 \centering
 \fbox{\includegraphics[width=0.7\textwidth]{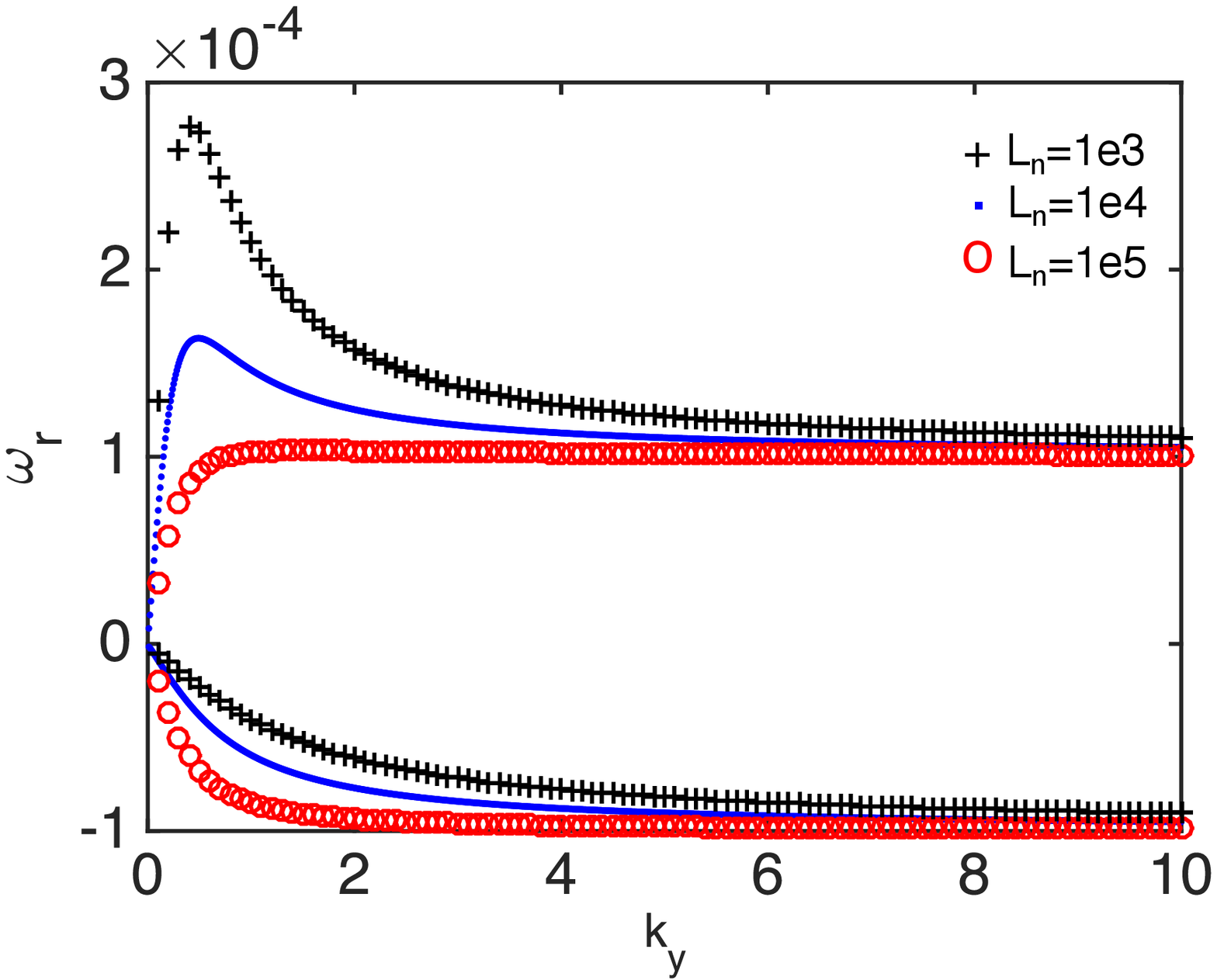}}
 \caption{The effect of gradient scale length $L_n$ on the dispersion relation of the drift wave coupled to dust acoustic wave ( $ b=n_{e0}/n_{i0}=0.15 $, $ k_z=10^{-6}$, $ \frac{C_D}{v_s} = 10^{-4} $)}
 \label{fig6}
 \end{figure}
 \end{center}
 The presence of dust dynamics has been found to modify the dispersion relation significantly. The Fig.(~\ref{fig3}) compares the dispersion relation of the drift wave proposed by Hasegawa-Mima (red color) with the drift wave coupled to the acoustic mode (blue color) when the dust dynamics has also been considered. The dispersion relation does not depend on the the value of $ k_z $ as shown in Fig.(~\ref{fig4}).  The mode frequency also depends on the value of depletion ratio $ \frac{n_{e0}}{n_{i0}} $ [Fig.(~\ref{fig5})]. The larger the depletion ratio, lesser is the  mode ferquency in the region where $k_y$ is smaller. The scale length of the density gradient plays a vital role in deciding the coupling of drift wave with dust acoustic wave. The sharp peak does not appears at all the values of the scale length $ L_n $ but emerges at a particular value of it. After this particular value, peak of the mode increases with increase in the length scale to a certain value as shown in Fig.(~\ref{fig6}). But the saturated mode which observed for the larger value of $ k_y $ does not changes with the change in the parameters $ k_z $, $ \frac{n_{e0}}{n_{i0}} $ and $ L_n $.

 \subsection{Strongly correlated dusty plasma ($ \mathbf{\omega \tau_m \gg 1} $)}
\begin{center}
  \begin{figure}[!h]
 \centering
 \fbox{\includegraphics[width=0.7\textwidth]{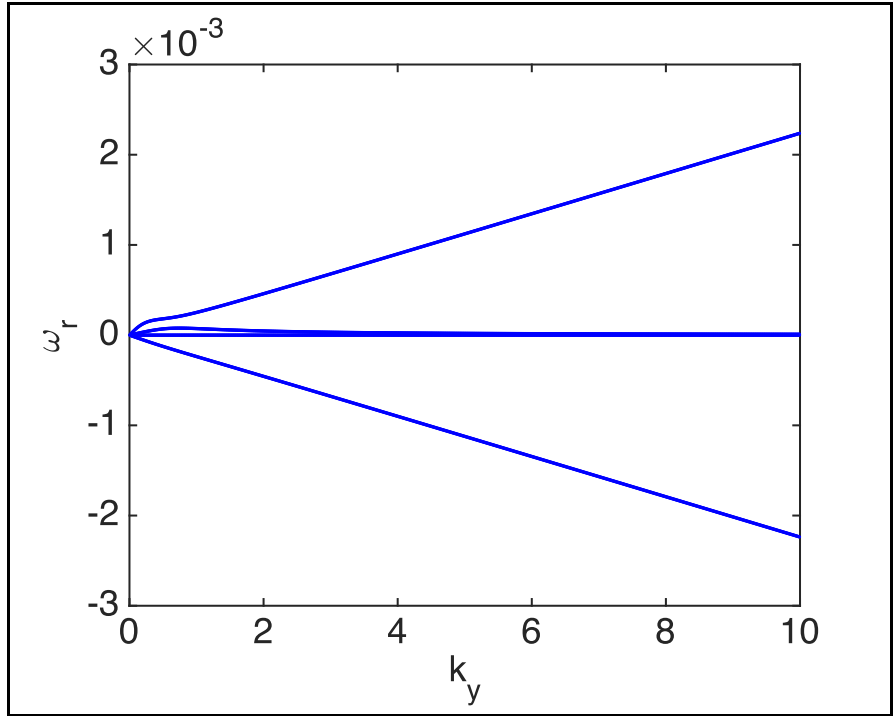}}
 \caption{ Dispersion relation of drift wave coupled to dust acoustic wave in strongly coupled dusty plasma  ($ a=\frac{\eta}{\tau_m}=10^{-8}$, $ k_z=10^{-6} $, $ L_n=10^4 $, $ \frac{C_D}{v_s} = 10^{-4} $, , $ \frac{n_{e0}}{n_{i0}}= 0.15 $)}
\label{fig7}
 \end{figure}
 \end{center}

\begin{center}
  \begin{figure}[!h]
 \centering
 \fbox{\includegraphics[width=0.7\textwidth]{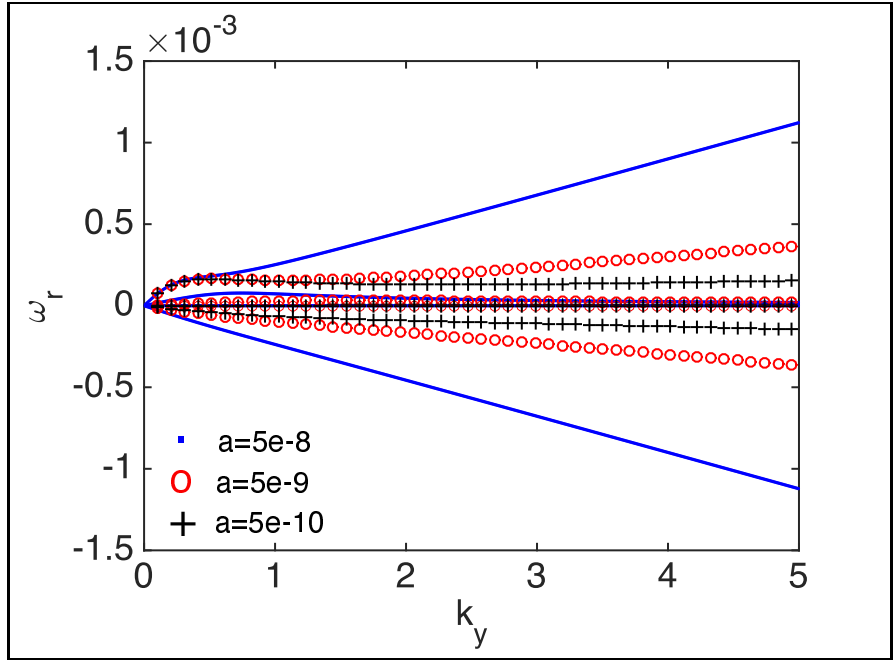}}
 \caption{ Dependence of  Dispersion relation of drift wave coupled to dust acoustic wave in strongly coupled dusty plasma on  $ a=\frac{\eta}{\tau_m}$ ($ k_z=10^{-6} $, $ \frac{C_D}{v_s} = 10^{-4} $, $ L_n=10^4$, $ \frac{n_{e0}}{n_{i0}}= 0.15 $)}
\label{fig8}
 \end{figure}
 \end{center}

 It evident from   (Fig.{~\ref{fig7}}) that for large value of $ k_y $, the frequency $\omega$ in strongly coupled regime vary as $ \omega \sim k_y $ which is nothing but the shear mode as the terms $ \bigg [ \frac{n_{e0}}{n_{i0}} + {k_\perp^2} \bigg ] {\omega}^4 $ and $ \bigg [ \bigg ( \frac{n_{e0}}{n_{i0}} + { k_{\perp}^2} \bigg ) \frac{\eta}{\tau _m} k^2 \bigg ]{ \omega}^2  $ balances each other which give rise to shear mode $ {\omega _r^2} = \frac{\eta}{\tau _m} k^2 $.There also exists a mode which is similar to the old drift wave mode which goes to zero for large value of $ k_y $. For large value of $ \frac{\eta}{\tau_m} $, the term $ \bigg [ \bigg ( \frac{n_{e0}}{n_{i0}} + { k_{\perp}^2} \bigg ) \frac{\eta}{\tau _m} k^2 \bigg ]{ \omega}^2 $ balances the term $  \frac{k_y}{L_n} \frac{\eta}{\tau _m} k^2 \omega $ which give rise to a modified drift mode $ \omega _r = \frac{k_y/L_n}{(n_{e0} / n_{i0} + {k_\perp^2})} $ which is deviated from that proposed by Hasegawa-Mima by a factor $ \frac{n_{e0}}{n_{i0}} $ in the denominator. Furthermore, for a small value of $ \frac{\eta}{\tau_m} $, the equation reduces back to what we obtained for the drift wave coupled to the dust acoustic wave in weakly coupled limit as expected as shown in Fig.(~\ref{fig8}).
 
 \begin{center}
  \begin{figure}[!h]
 \centering
 \fbox{\includegraphics[width=0.7\textwidth]{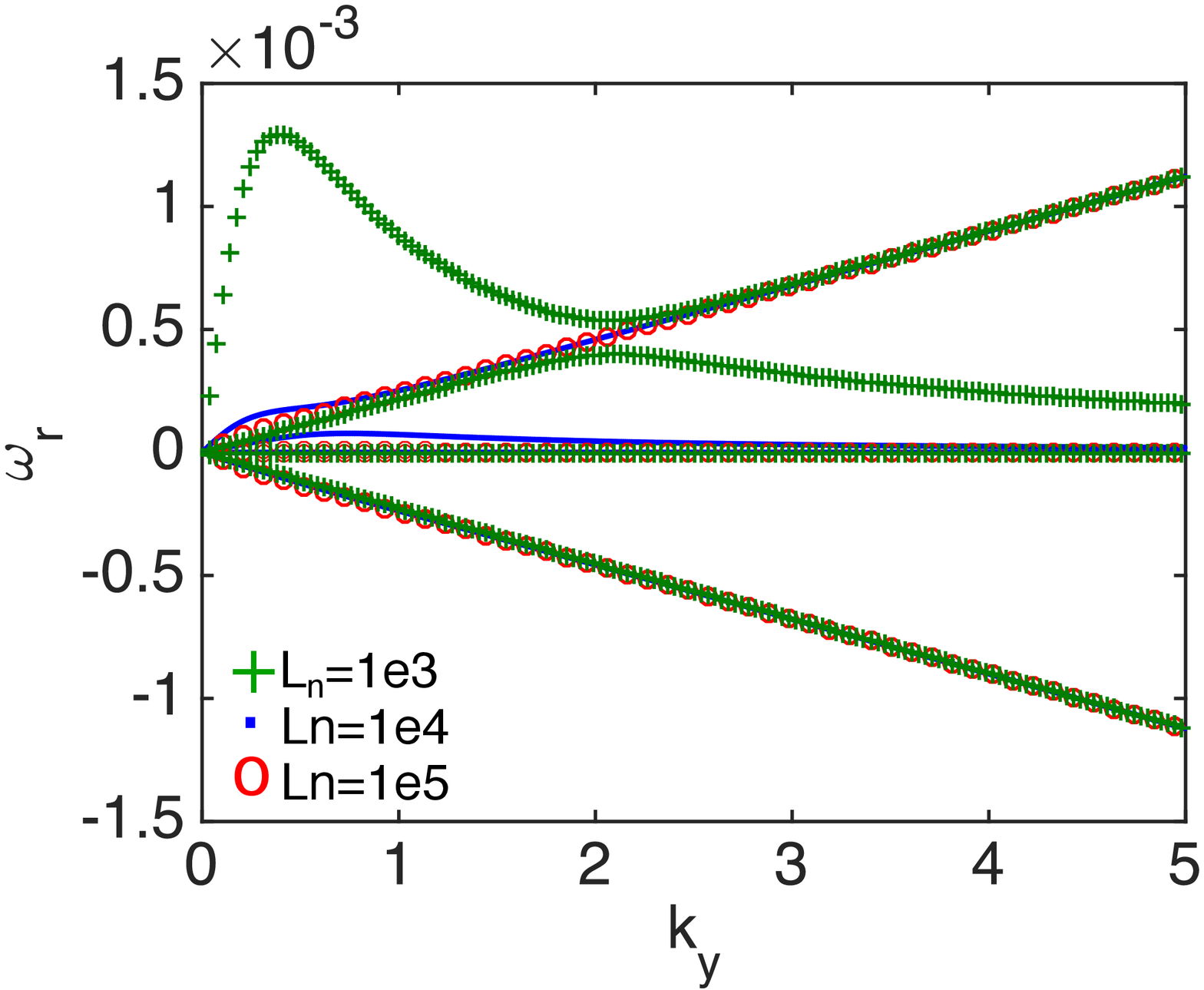}}
 \caption{ Dependence of dispersion relation on $ L_n$( $ k_z=10^(-6) $, $ \frac{C_D}{v_s} = 10^{-4} $, $ \frac{\eta}{\tau_m}= 1 \times 10^{-8}$)}
\label{fig9}
 \end{figure}
 \end{center}
 \begin{center}
  \begin{figure}[!h]
 \centering
 \fbox{\includegraphics[width=0.7\textwidth]{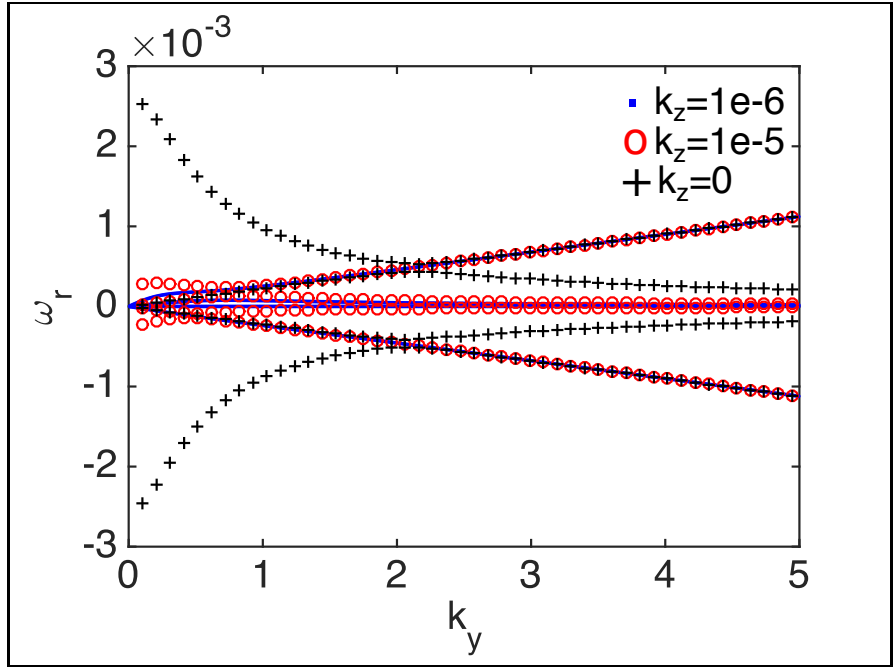}}
 \caption{ Dependence of dispersion relation on gradient scale length $k_z$ ($ L_n=10^{4} $, $ \frac{C_D}{v_s} = 10^{-4} $, $ \frac{\eta}{\tau_m}= 1\times 10^{-8}$, $ \frac{n_{e0}}{n_{i0}}= 0.15 $)}
\label{fig10}
 \end{figure}
 \end{center}
 
 \begin{center}
  \begin{figure}[!h]
 \centering
 \fbox{\includegraphics[width=0.7\textwidth]{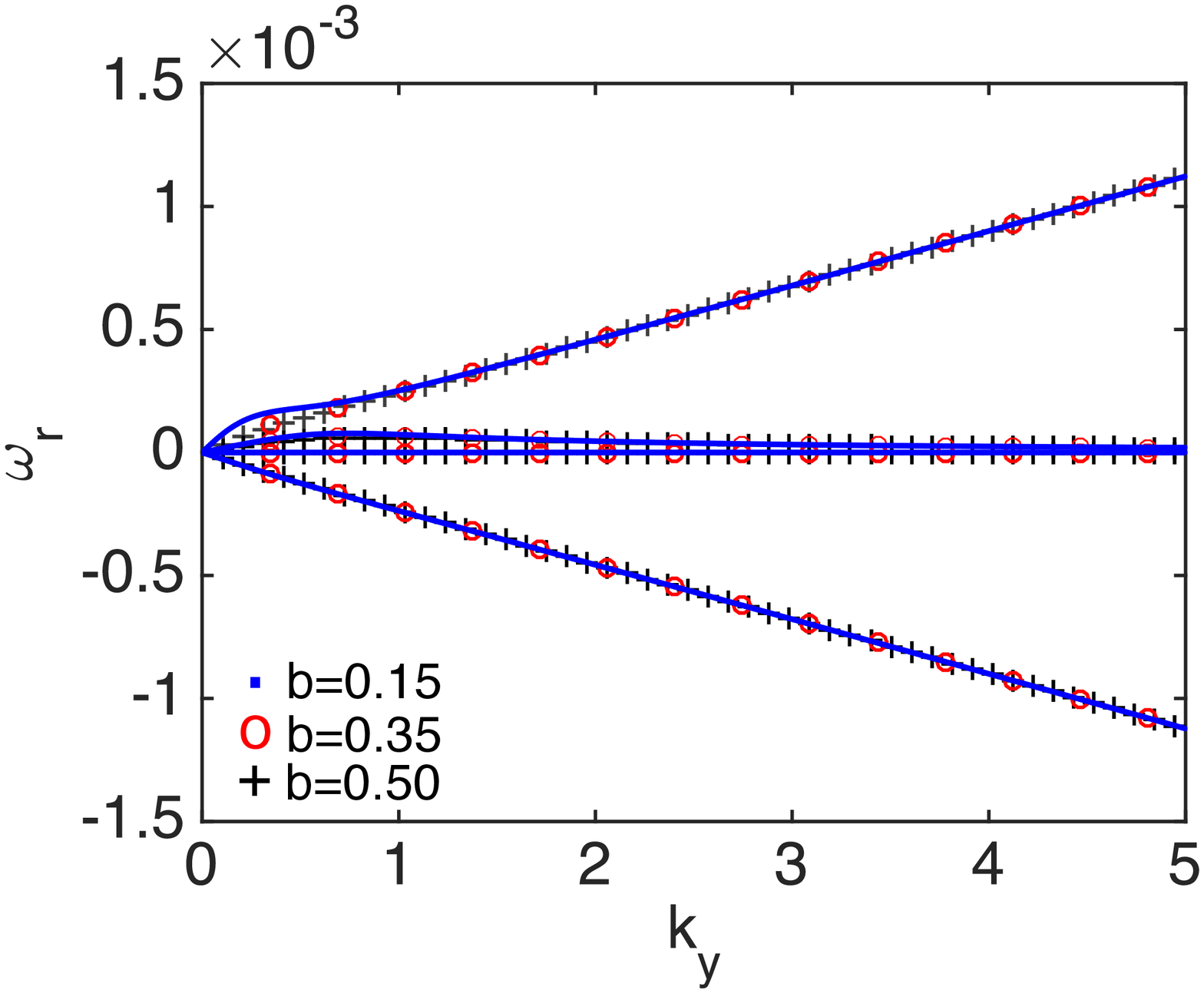}}
 \caption{ Dependence of dispersion relation on $ b=n_{e0}/n_{i0}$( $ L_n=10^4 $, $ \frac{C_D}{v_s} = 10^{-4} $, $ \frac{\eta}{\tau_m}= 1 \times 10^{-8}$, $k_z=10^{-6}$)}
\label{fig11}
 \end{figure}
 \end{center}
We do not observe any  growing mode in the strongly coupled regime (Fig.8). The dependence of dispersion relation on the parameters $ \frac{\eta}{\tau_m} $, gradient scale length $ L_n $,  $ k_z $ and $b=n_{e0}/n_{i0}$ has been shown in figures (8), (9), (10) and (11) respectively.

\section{ summary and conclusions}
In weakly coupled nonuniform, magnetized dusty plasma, when the dust dynamics are also considered, there exists a possibility of coupling of drift wave with the dust acoustic wave in strong magnetic field approximation ( $ \omega \ll \omega_{ci} $ ). Now the drift mode has been significantly modified as compared to the earlier Hasegawa-Mima drift wave in the presence of the dust dynamics. For the large value of $ k_y $, we observe a saturated mode which is independent of $ k_\bot $. In this region, shielding is done by the polarization drift in the perpendicular direction. In addition, the dispersion properties of the drift mode is strongly modified with the parallel wave vector $ k_z $, the factor $ \frac{n_{e0}}{n_{i0}} $ and  the length scale of the density gradient $ L_n $. However,  saturated mode obtained for  larger value of $ k_y $ does not changes with the change in the parameters $ k_z $, $ \frac{n_{e0}}{n_{i0}} $ and $ L_n $.

In strongly coupled nonuniform magnetized dusty plasma,  fourth order in $ \omega $ for drift wave coupled with dust acoustic wave in the kinetic regime ( $ \omega \tau_m \gg 1 $ ) has been obtained. Drift mode in strongly coupled regime converts to transverse shear mode at large  $ k_y $. For smaller $ a = \frac{\eta}{\tau _m} $, the equation reduces back to that obtained in weakly coupled limit. The  dust shear mode coupled to ion drift mode has been found to be the natural mode of plasma.

In conclusion, the dust dynamics (in both weakly and strongly correlated) has been shown to  significantly influence the ion drift in an inhomogeneous, magnetized plasma and can be observed in laboratory experiments as well as in astrophysical plasma.


\bibliographystyle{unsrt}

\begin{thebibliography}{999}
\addcontentsline{toc}{chapter}{\numberline{}Bibliography}
 

 
 \bibitem{BB:1993}
  Rao, N. ~(1993),  
{\it Low-frequency waves in magnetized dusty plasmas}, 
 Journal of Plasma Physics,
 {\bf 49,3} 
 
 \bibitem{BB:1996}
   Praburam G. and  Goree J. ~(1996),  
{\it Experimental observation of very low‐frequency macroscopic modes in a dusty plasma}, 
 Physics of Plasmas ,
 {\bf 3,1212} 
 
 
 
 \bibitem{BB:19961}
 Barkan A., D'Angelo N. and  Merlino R.L.~(1996),  
{\it Experiments on ion-acoustic waves in dusty plasmas}, 
 Planetary and Space Science,
 {\bf 44,3} 
 



\bibitem{BB:68}
Krall N. A.~(1968),  
{\it Advances in Plasma Physics vol 1 ed A Simon and W B Thompson}, 
Interscience, New York 
 pp~153.

\bibitem{BB:77}
Hasegawa, Akira and Mima, Kunioki~(1977),  
{\it Stationary Spectrum of Strong Turbulence in Magnetized Nonuniform Plasma}, 
Phys. Rev. Lett., 
{\bf 39,4}, pp~205-208.

\bibitem{BB:78}
Hasegawa, Akira and Mima, Kunioki~(1978),  
{\it Pseudo‐three‐dimensional turbulence in magnetized nonuniform plasma}, 
Physics of Fluids (1958-1988), 
{\bf 21,1}, pp~87-92.

\bibitem{BB:1999}
Horton, W.~(1999),  
{\it Drift waves and transport}, 
 Rev. Mod. Phys.,
 {\bf 71, 735} 

\bibitem{BB:01}
Shukla, P. K.~(2001),
{A survey of dusty plasma physics},
Physics of Plasmas (1994-present),
{\bf 8,5}, pp~1791-1803

\bibitem{BB:02}
Shukla, P. K.,Mamun, A. A. ~(2002),
{\it Introduction to Dusty Plasma}, 
Inst. of Physics, Bristol


\bibitem{BB:09}
Shukla, P. K. and Eliasson, B.~(2009),
{\it Colloquium}, 
Rev. Mod. Phys
{ \bf 81,1 }, pp~25-44. 

\bibitem{BB:931}
Shukla, P. K. and Varma, R. K.~(1993),
{\it Convective cells in nonuniform dusty plasmas},
Physics of Fluids B: Plasma Physics (1989-1993),
{\bf 5,1}, pp~236-237.



\bibitem{BB:991}
Pokhotelov O. A., Onishchenko O. G., and Shukla, P. K.~(1999),
{\it Drift-Alfven vortices in dusty plasmas },
Journal of Geophysical Research,
{\bf 104, A9}, pp~385-389.

\bibitem{BB:031}
Haque, Q. and Saleem, H.~(2003),
{\it Nonlinear dynamics of electrostatic and electromagnetic drift modes in dusty plasmas}
Journal of Geophysical Research: Space Physics,
{\bf 108,A12}, pp~2156-2202.







\bibitem{BB:962}
Rosenberg, M. and Krall, N. A.~(1996),
{\it Low frequency drift instabilities in a dusty plasma},
Physics of Plasmas (1994-present),
{\bf 3,2}, pp~644-649.

\bibitem{BB:2015}
Thomas, E. \emph{et. al.}~(2015),
{\it The magnetized dusty plasma experiment (MDPX)},
Journal of Plasma Physics,
{\bf 81(2)}.

\bibitem{BB:2012}
 Thomas E. Jr.,  Merlino r. L. and  Rosenberg M.~(2012),
{\it Magnetized dusty plasmas: the next frontier for complex plasma research},
Plasma Physics and Controlled Fusion,
{\bf 54, 12}.

\bibitem{BB:86}
Ikezi, H. (1986)
{\it Coulomb solid of small particles in plasmas}
Physics of Fluids (1958-1988)
{\bf 29,6}, pp~1764-1766.

\bibitem{BB:04PRL}
Oliver Arp, Dietmar Block, and Alexander Piel (2004)
{\it Dust Coulomb Balls: Three-Dimensional Plasma Crystals}
Phys. Rev. Lett.,
{\bf 93,16}.

\bibitem{BB:94PRL}
H. Thomas, * G. E. Morfill, V. Demmel and J. Goree (1999)
{\it Plasma Crystal: Coulomb Crystallization in a Dusty Plasma}
Phys. Rev. Lett.,
{\bf 73,5}.

\bibitem{BB:921}
Berkovsky, M. A.~(1992),
{\it Spectrum of low frequency modes in strongly coupled plasmas},
Physics Letters A,
{\bf 166,5-6}, pp~365-368.

\bibitem{BB:98}
Kaw, P. K. and Sen, A.(1998),  
{\it Low frequency modes in strongly coupled dusty plasmas}, 
Physics of Plasmas (1994-present) 
{\bf 5,10}, pp~3552-3559.

\bibitem{BB:142}
 Das, A., Dharodi, V. and Tiwari S.~(2014),
{\it Collective dynamics in strongly coupled dusty plasma medium}, 
Journal of Plasma Physics,
{\bf 80}, pp~855-861.

\bibitem{BB:14}
Dharodi, V. S. and Tiwari, S. K. and Das, A.~(2014),
{\it Visco-elastic fluid simulations of coherent structures in strongly coupled dusty plasma medium},
Physics of Plasmas,
{\bf 21,7}.

\bibitem{BB:071}
Bandyopadhyay, P. and Prasad, G. and Sen, A. and Kaw, P.K..~(2007),
{\it Experimental observation of strong coupling effects on the dispersion of dust acoustic waves in a plasma},
Physics Letters A,
{\bf 368,6}, pp~491-494.






\






\end{thebibliography}

\end{document}